\documentclass[aps,prl,preprint,nopacs,amsmath,superscriptaddress]{revtex4}
\usepackage{epsf}
\usepackage{graphicx}
\usepackage{sidecap}
\usepackage{array}
\usepackage{amsmath}
\usepackage{amssymb}
\usepackage{dcolumn}
\usepackage{epstopdf}
\usepackage{bm}% bold math
\usepackage{color}

\def \beq {\begin{equation}}
\def \eeq {\end{equation}}
\begin{document}
\title{{Observation of Gapless Dirac Surface States in ZrGeTe}}
%\title{{Observation of a topological insulating surface state in the non-symmorphic Dirac semimetal ZrGeTe}}

\author{M.~Mofazzel~Hosen}\affiliation {Department of Physics, University of Central Florida, Orlando, Florida 32816, USA}
\author{Klauss~Dimitri}\affiliation {Department of Physics, University of Central Florida, Orlando, Florida 32816, USA}
\author{Alex Aperis}
\affiliation {Department of Physics and Astronomy, Uppsala University, P.\,O.\ Box 516, S-75120 Uppsala, Sweden}
\author{Pablo Maldonado}
\affiliation {Department of Physics and Astronomy, Uppsala University, P.\,O.\ Box 516, S-75120 Uppsala, Sweden}
\author{Ilya~Belopolski}
\affiliation {Joseph Henry Laboratory and Department of Physics, Princeton University, Princeton, New Jersey 08544, USA}
\author{Gyanendra Dhakal} 
\affiliation {Department of Physics, University of Central Florida, Orlando, Florida 32816, USA}
\author{Firoza Kabir}
\affiliation {Department of Physics, University of Central Florida, Orlando, Florida 32816, USA}
\author{Christopher~Sims}
\affiliation {Department of Physics, University of Central Florida, Orlando, Florida 32816, USA}
\author{M.~Zahid~Hasan}
\affiliation {Joseph Henry Laboratory and Department of Physics,
	Princeton University, Princeton, New Jersey 08544, USA}
\author{Dariusz Kaczorowski}
\affiliation {Institute of Low Temperature and Structure Research, Polish Academy of Sciences,
	50-950 Wroclaw, Poland}
\author{Tomasz~Durakiewicz}
\affiliation {Condensed Matter and Magnet Science Group, Los Alamos National Laboratory, Los Alamos, NM 87545, USA} 
\affiliation {Institute of Physics, Maria Curie - Sklodowska University, 20-031 Lublin, Poland}
\author{Peter M.\ Oppeneer}
\affiliation {Department of Physics and Astronomy, Uppsala University, P.\,O.\ Box 516, S-75120 Uppsala, Sweden}
\author{Madhab~Neupane*}
\affiliation {Department of Physics, University of Central Florida, Orlando, Florida 32816, USA}

\date{\today}
%\pacs{}
\begin{abstract}
\noindent
{The experimental discovery of the topological Dirac semimetal establishes a platform to search for various exotic quantum phases in real materials. 
ZrSiS-type materials have recently emerged as topological nodal-line semimetals where gapped Dirac-like surface states are observed. 
Here, we present a systematic angle-resolved photoemission spectroscopy (ARPES) study of ZrGeTe, a non-symmorphic symmetry protected Dirac semimetal.
%Our experimental data reveal the presence of multiple Fermi surface pockets, but
%such as the diamond-shaped Fermi surface at the zone center ($\Gamma$), elliptical-shaped Fermi surface at the M point, and a small electron pocket at the X point of the BZ, respectively. 
 We observe two Dirac-like gapless surface states at the same $\bar{X}$ point of the Brillouin zone.
Our theoretical analysis  and first-principles calculations reveal that these are protected by crystalline symmetry. 
%Our systematic studies reveal that 
Hence, ZrGeTe appears as a rare example of a naturally fine tuned system where the interplay between symmorphic and non-symmorphic symmetry leads to rich phenomenology, and thus opens for opportunities to investigate the physics of Dirac semimetallic and topological insulating phases realized in a single material.}
%provide a platform to engineer the two dimensional Dirac fermion state.}

\end{abstract}

\date{\today}
\maketitle

%Introduction:

%The experimental discovery of the topological insulator (TI) \cite{Hasan, SCZhang, Hasan_review_2, Xia, Neupane_4} tremendously motivated research in topological Dirac semimetals (TDS) which offers a platform for realizing many exotic physical phenomena such as high mobility electron, magnetoresistance, etc. The TDSs \cite{Dai, Neupane, Neupane_2, Cava1, Nagaosa, Young3D, Young_Kane, Xi_Dai, Chen, Neupane_6, NdSb, Zhang} are characterized by symmetry protected band touchings at certain \textit{k}-points of the Brillouin zone (BZ).
%In addition to 0D band touchings, topological nodal-line semimetals form a 1D line or loop in k-space \cite{node_0, node_1, node_2, node_3, node_4, Dai_LiFeAs, Neupane_5, Schoop, Ding, tranp1, PTS, PT, new_ZST, HfSiS, ZrSiX} which requires extra symmetry to be protected aside from the translational symmetry. Importantly, this point or line contact can be lifted by tuning the spin-orbit coupling (SOC) strength .
The experimental discovery of the topological insulator (TI) \cite{Hasan, SCZhang, Hasan_review_2, Hasan_2, Xia, Neupane_4} tremendously motivated research in topological Dirac semimetals (TDS) which offer a platform for realizing many exotic physical phenomena such as high electron mobility and magnetoresistance. The TDSs \cite{Hasan_2, Dai, Neupane, Neupane_2, Cava1, Nagaosa, Young3D, Young_Kane, Xi_Dai, Chen, Neupane_6, NdSb, Zhang} are characterized by symmetry protected band touchings at certain \textit{k}-points of the Brillouin zone (BZ), 1D lines or loops (topological nodal-line semimetals) \cite{node_0, node_1, node_2, node_3, node_4, Neupane_5, Schoop, Ding, tranp1, PTS, PT, new_ZST,ZrSiTe, HfSiS, ZrGeM, MSiS,ZrSiX, CeSbTe, GdSbTe}. Importantly, this point or line contact can be lifted by tuning the spin-orbit coupling (SOC) strength \cite{ZrGeM, ZrSiX, tune}.

A distinct class of TDS is characterized by the presence 
%and protection 
of non-symmorphic symmetries \cite{Young_Kane,Schoop, Neupane_5}. Such non-symmorphic materials possess forced band degeneracies at high symmetry points of the BZ as a consequence of fractional translational symmetry combined with a point group symmetry such as mirror or rotational symmetry. In the presence of time reversal symmetry, the Dirac points are pinned at the time reversal invariant momenta of the BZ \cite{Young_Kane}. This forced band degeneracy cannot be lifted by any perturbation including spin-orbit coupling. %Such robust bulk Dirac points have been predicted both for 3D \cite{Young3D} and 2D \cite{Young_Kane} systems but yet only observed in 3D \cite{Schoop,Neupane_5,ZrSiTe}. A non-symmorphic symmetry protected nodal line phase has been reported in ZrSiS-type materials \cite{ZrSiTe, MSiS} together with the presence of gapped Dirac cone surface state.
Such robust bulk Dirac points have been observed in 3D systems while a non-symmorphic symmetry protected nodal line phase has been reported in ZrSiS-type materials \cite{Neupane_5, Schoop, Ding, PTS, PT, new_ZST, HfSiS, ZrGeM, ZrSiX,MSiS,ZrSiTe} with the presence of a gapped Dirac cone surface state.

\begin{figure*}
	\centering
	\includegraphics[width=13.7cm]{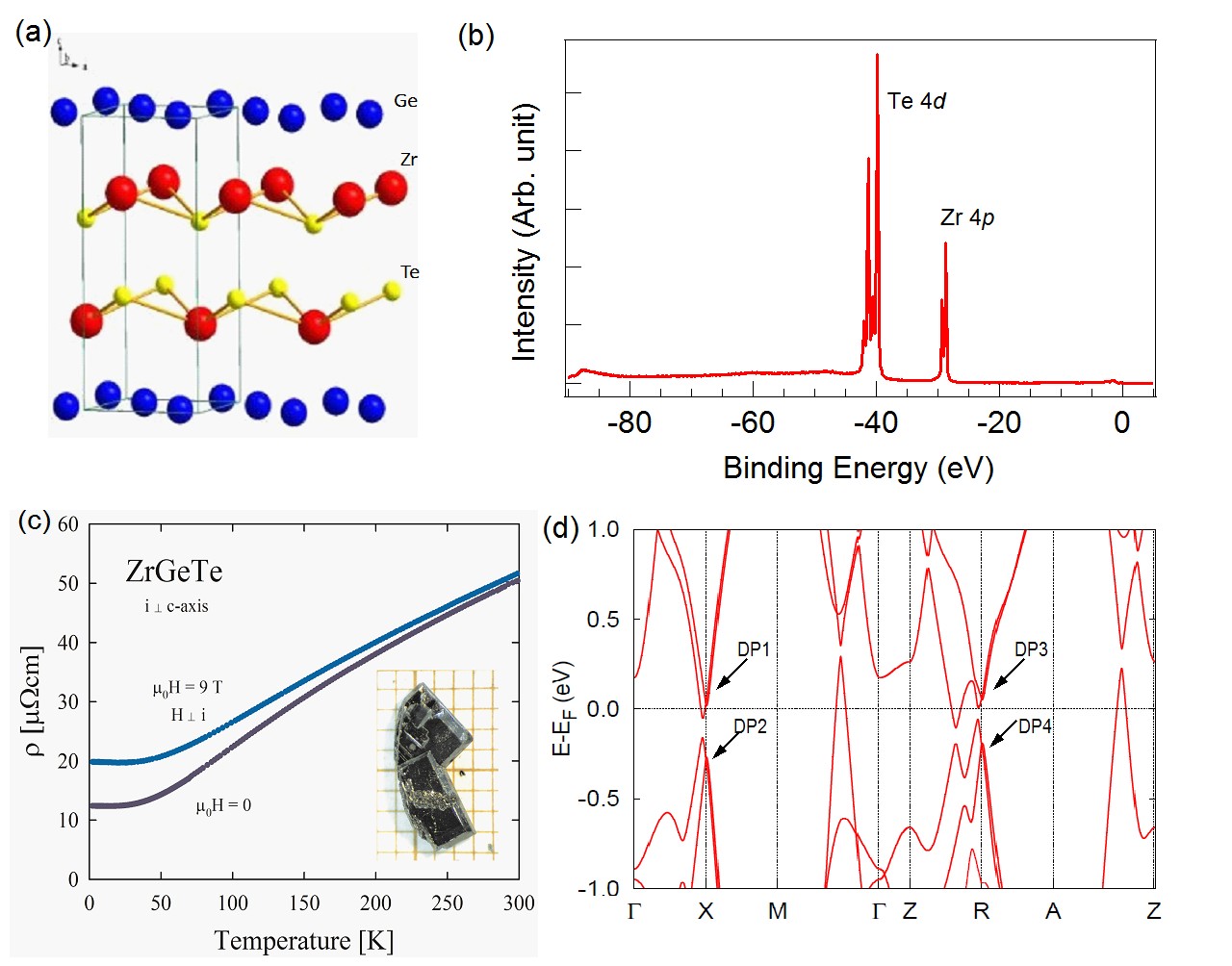}
	\caption{{Crystal structure and sample characterization of ZrGeTe}.  {(a)} Tetragonal crystal structure of ZrGeTe. The Zr layers are separated by two neighboring Te layers, which are both sandwiched between the Ge atoms forming a square net. 
		{(b)} Spectroscopically measured core-level of ZrGeTe. Sharp peaks of Te 4\textit{d} and Zr 4\textit{p} are observed.
		{(c)} Temperature variation of the electrical resistivity of ZrGeTe within the tetragonal plane measured in zero magnetic field and in a field of 9 T applied along the c-axis. Inset shows a single crystal of ZrGeTe.
		(d) \textit{Ab initio} calculated band structures of ZrGeTe along 
		%the various 
		high symmetry directions. Bulk Dirac points (DP) are indicated in the plot.}
\end{figure*}

Being essential for TDS, the space group symmetry protection plays a pivotal role in topological insulators (TIs) as well. While the so-called non-symmorphic TIs \cite{FangFu,Shiozaki,hourglass} are currently sought after, their symmorphic counterparts i.e.\ the topological crystalline insulators (TCIs) are already well established \cite{TCI3,TCI4,TCI5} after they were first theoretically predicted \cite{TCI}. A striking feature of TCIs is that they can host band inversions at multiple non-equivalent time-reversal invariant momenta (TRIMs) of their bulk BZ, depending on the orientation of the cleaved surface, may project onto the same TRIM of the surface BZ \cite{TCI2}. If the number of such projected points is even, then one would naively expect a trivial insulating phase since the overall $Z_2$ \cite{FuKanePRB} indices would vanish. However, if the cleaved surface preserves a crystalline symmetry of the bulk such as a rotation or mirror symmetries, then this extra symmetry can protect the surface states, thus giving rise to the TCI phase. Interestingly, such gapless surface states may exhibit multifold, anisotropic Weyl-like dispersions which are more complex than the usual strong or weak TI surface states \cite{TCI2}.

So far, little is known about the interplay of TDS and TI phases and whether these two phases can coexist in the same non-symmorphic material. 
In this letter, we report a systematic investigation of ZrGeTe, a member of the ZrSiS-type material family, combining both ARPES measurements and first-principles calculations, 
and establish ZrGeTe as a new material with multiple gapless Dirac fermionic surface states, reminiscent of that encountered in TCIs. We show that in contrast to typical TCIs, the gapless surface states in ZrGeTe stem from a double band inversion that takes place at a single non-equivalent bulk TRIM, due to an inversion between two non-symmorphic symmetry protected bulk Dirac points. At the surface, the residual symmorphic part of this symmetry remains and protects the resulting surface states. Our findings suggest that ZrGeTe can aptly function as a new platform for studying exotic properties caused by coexistence of TDS and TI phases in one material, due to the interplay between symmorphic and non-symmorphic symmetry. 

Single crystals of ZrGeTe were grown by chemical vapor transport method using iodine as transporting agent \cite{SI}. Similar to the MSiX, where M = Zr, Hf and X = S, Se, Te, respectively, it also crystallizes in the PbFCl-type crystal structure with space group $P4/nmn$ (No.\ 129). Fig.\ 1(a) illustrates the layered \cite{holy} crystal structure of ZrGeTe. The Zr atom layers are separated by two Te layers and the Zr-Te terminals are sandwiched between a Ge square net. 
%The larger size of the Ge atom is responsible for the increased c/a ratio compared to the other MSiX materials. 
The bond between Zr-Te is relatively weak compared to the bond between Zr-Ge hence providing a natural cleaving plane between the two neighboring Zr-Te layers and therefore cleaves easily along the (001) plane. Furthermore, the square Si-net provides a natural glide plane and the crystal preserves two sets of non-symmorphic symmetry. Fig.\ 1(b) presents the spectroscopic core level measurement of ZrGeTe. We observe the sharp peaks of Zr 4\textit{p} ($\sim$28.5 eV) and Te 4\textit{d} ($\sim$40 eV) of ZrGeTe, which further indicate the excellent quality of the samples used for our measurements. 
As shown in Fig.\ 1(c), ZrGeTe exhibits metallic-like electrical conductivity, with the temperature dependence similar to that reported for ZrSiTe \cite{ZrSiX}. At room temperature, the resistivity measured within the basal plane of the tetragonal unit cell (i $\perp$ c-axis) amounts to 50 $\mu\Omega$cm. With decreasing temperature, it decreases smoothly down to about 12 $\mu\Omega$cm at 2 K. In magnetic field of 9 T, applied along the c-axis, the overall character of $\rho$(T) hardly changes, in striking contrast to the behavior in ZrSiS and ZrSiSe \cite{Zhang, ZrSiX}, yet fairly reminiscent to the case of ZrSiTe \cite{ZrSiX}. At 2 K, the transverse (H $\perp$ i) magnetoresistance of ZrGeTe, defined as MR = [$\rho$(T, $\mu_0$H = 9 T) - $\rho$(T,0]/$\rho$(T,0), is positive and attains about 60$\%$. It is twice larger than the MR in ZrSiTe \cite{ZrSiX}, yet a few orders of magnitude smaller than the other Si-bearing counterparts \cite{Zhang, ZrSiX}. 
Fig.\ 1(d) presents the calculated bulk bands along various high symmetry directions.

\begin{figure*}
	\centering
	\includegraphics[width=14.85cm]{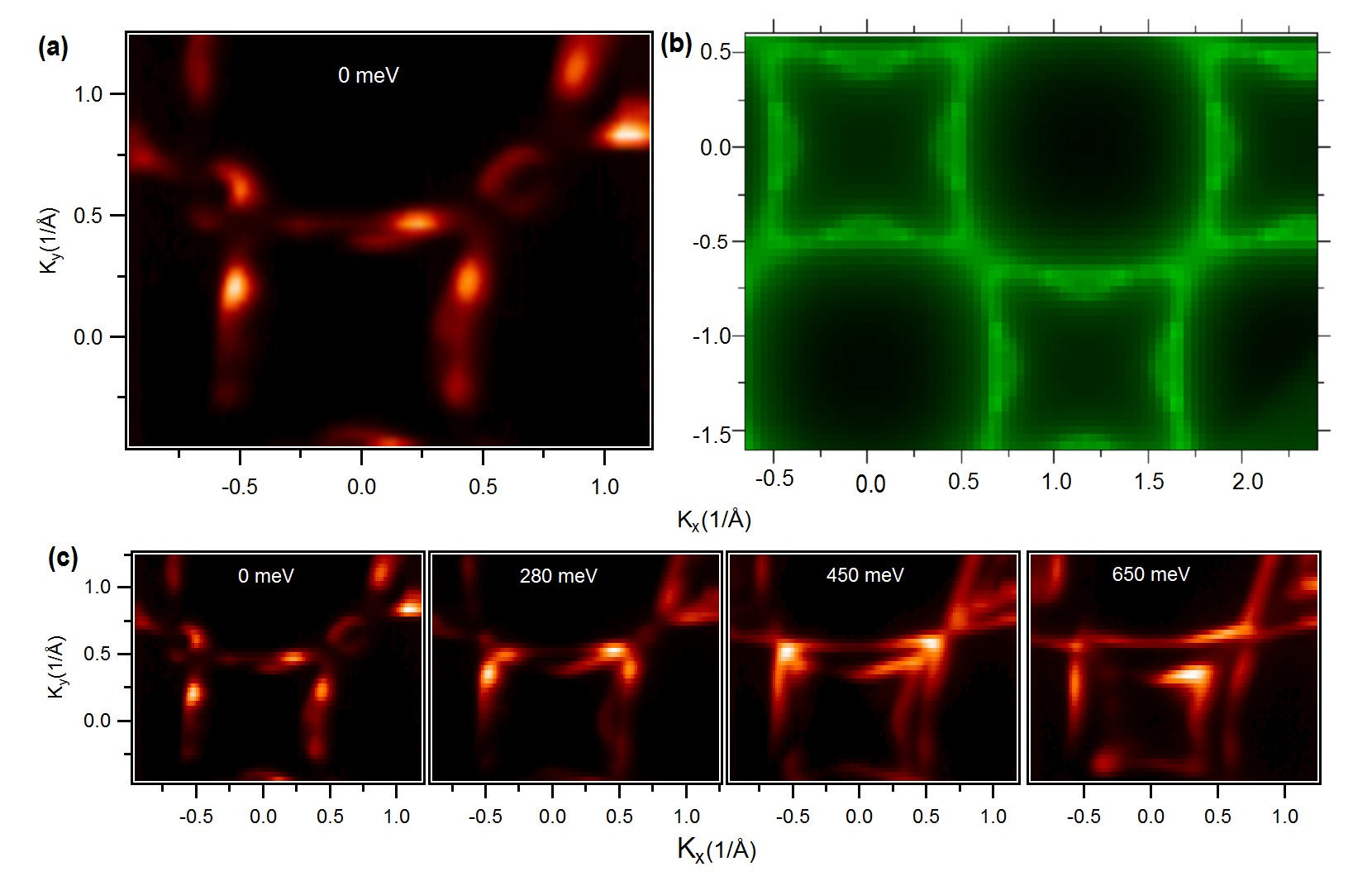}
	\caption{{Fermi surface map and constant energy contours of ZrGeTe.} {(a)} Fermi surface map measured with a photon energy of 60 eV. {(b)} \textit{Ab initio} calculated Fermi surface map. {(c)} Measured Fermi surface and constant energy contours. The binding energy values are noted in the plots. All measurements were performed at the ALS beamline 10.0.1 at a temperature of $\sim$15\,K. }
\end{figure*}

%In order 
To unveil the detailed electronic structure of ZrGeTe, we present our ARPES measured electronic structure in Figs.\ 2 and 3. Fig.\ 2(a) shows the Fermi surface map measured at 
%the ALS beamline 10.0.1 at 
a photon energy of 60\,eV and a temperature of 15\,K. Experimentally we observed a diamond-shaped Fermi surface around the $\Gamma$ point, circular-shaped Fermi pockets at the X point, and a nearly ellipsoidal-shaped Fermi surface around the M points of the BZ. Our measured Fermi surface shows excellent agreement with our calculated bulk Fermi surface presented in Fig.\ 2(b). Moving toward higher binding energies the diamond shaped Fermi surface becomes disconnected resembling an inner diamond within an outer diamond (see Fig.\ 2(c). The two diamond-shaped Fermi surfaces are clearly disconnected along the M-$\Gamma$-M direction. At around 450 meV the circular-shaped Fermi pockets shrink into a point-like state located around the binding energy of the Dirac points. This reveals the electron like nature of the bands around the X points of the BZ. Moreover, our first-principles calculations identify the surface origin of the bands around the X points (see below) which is also consistent with previously reported data on other materials of this family \cite{Neupane_4, Schoop,MSiS, ZrSiTe,HfSiS}.

\begin{figure*}
	\centering
	\includegraphics[width=0.8\linewidth]{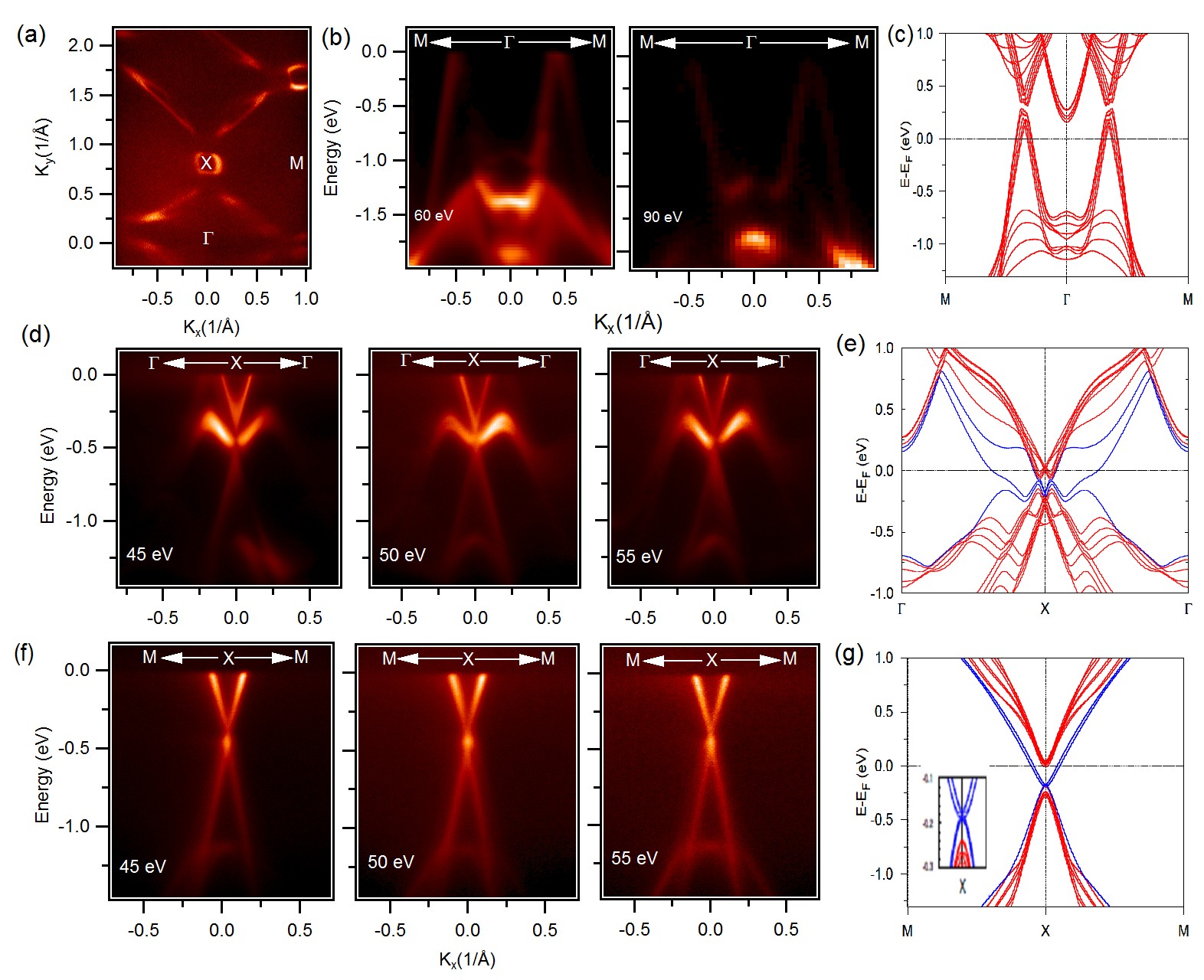}
	\caption{{Observation of gapless Dirac fermion state.} {(a)} Measured Fermi surface at a photon energy of 40 eV. High-symmetry points are indicated in the plot. {(b)} Measured dispersion maps along the M-$\Gamma$-M direction at photon energies of 60 eV and 90 eV, respectively. {(c)} \textit{Ab initio} calculations showing the dispersion map along the M-$\Gamma$-M high-symmetry direction. {(d)} Photon energy dependent ARPES dispersion maps along the $\Gamma$-X-$\Gamma$ direction. The photon energies are noted on the plots. \textbf{(e)} Calculated dispersion map along the high-symmetry direction $\Gamma$-X-$\Gamma$. {(f)} ARPES energy-momentum dispersion maps measured with various photon energies along the M-X-M direction. Linearly dispersive states are observed which do not show 
		dispersion with a dependence on photon energy. 
		The photon energies are noted on the plots. (g) Calculations showing the dispersion map along the M-X-M high-symmetry direction. The inset show a zoomed in plot of bands around the X point which clearly show the band touching, in qualitative agreement with the experimental data. All experiments were performed at the ALS end-stations 10.0.1 and 4.0.3 at a temperature of $\sim$$15-20$\,K.}
\end{figure*}

%\bigskip
%\bigskip
%\textbf{Results}

%Now we discuss the electronic band structure of the ZrGeTe system. The calculated bulk band structure along various high-symmetry directions is shown in Fig. 1d. Small-gaps are observed due to the spin-orbit coupling. The measured Fermi surface maps using photon energies of 70 eV and 30 eV are shown in Fig. 1e and Fig. 1f, which are obtained within the energy window of $\pm$ 5 meV near the Fermi level. Various Fermi pockets are observed at the Fermi surface map obtained at 70 eV by covering a larger area of the Brillouin zone. Specifically, a diamond shaped Fermi surface is observed around the zone center ($\Gamma$) point, and an ellipsoidal pocket is seen at around the M point. Interestingly, we also observe a small electron-like Fermi pocket around the X point of the BZ. Fig. 1g shows the calculated bulk Fermi surface map. Comparing  experimental Fermi surface maps in Fig. 1e-f with the calculated Fermi surface shown in Fig. 1g, it is clear that the states around the X point are not originated from the bulk band. We attribute these states to the surface electronic structure, in good agreement with  the slab calculations. 

To reveal the nature of the states around the X point and along various high-symmetry directions, it is necessary to perform photon energy dependent ARPES measurements as shown in Fig.\ 3. Fig.\ 3(a) shows the Fermi surface map with the high-symmetry points are noted in the plot. 
%The Fermi surface measurements were performed at a photon energy of 40 eV and at a temperature of 15 K. 
Fig.\ 3(b) shows the photon energy dependent dispersion maps along the M-$\Gamma$-M direction. It shows good qualitative agreement with our calculated dispersion map as shown in Fig.\ 3(c). Our calculations predicted a nodal-line state along the M-$\Gamma$-M direction, however, it is located above the Fermi level. Fig.\ 3(d) shows the dispersion maps along the $\Gamma$-X-$\Gamma$ direction at various photon energies. By comparing with the slab calculations along this direction [see Fig.\ 3(e)], we can clearly identify the  surface [blue color in Fig.\ 3(e,g)] and bulk [red color in Fig.\ 3(e,g)] originated states. {The upper V shape in the experimental data clearly originates from the surface state, however, in the $\mathcal{M}$ shaped states, the bulk states interfere with the surface states.} 
%{\blue An important point to notice here is that our calculation predicted the forced gap closing as a result of the non-symmorphic symmetry protection at the X point along the M-X-M direction which is consistent with our experimental observations.} 
An important point to notice here is that our calculations predicted a gapless Dirac surface state at the X point over a wide energy range along the M-X-M direction, which is consistent with our experimental observations.
To further confirm this, we have performed photon energy dependent dispersion mapping along this direction. Fig.\ 3(f) represents the dispersion maps along the M-X-M high symmetry direction at different photon energies, as noted on the plots. The Dirac points are observed to be located at around 400 meV and do not disperse with photon energy.  {We note the presence of bright intensity in the vicinity of the Dirac point in Fig.\ 3(f). As we discuss further below, this feature is due to the interference between the near degenerate Weyl states that form the gapless surface states.}
%One possible reason of this feature could be due to  interference of the bulk Dirac points (see DP2 and DP4 in Fig. 1d), since those are situated slightly below the surface state crossing. 
The experimentally observed states %presented in Fig.\ 3(f), 
are reproduced nicely in our calculations. The surface states do not show any gap as compared to the bulk states [see Fig.\ 3(f)] which is also consistent with our band calculations (see Fig.\ 3(e,g) and SM). The presence of eight surface states in total in Fig. 3(g) are due to the fact that in the slab calculations there is a top and a bottom surface contributing.

Our calculation reveals that ZrGeTe is a Dirac semimetal with bulk Dirac crossings at X and R, in accordance with other members of this family \cite{ZrSiTe}. This picture is consistent with our bulk calculations where these crossings are found to occur near the Fermi level at energies around 83 meV and -200 meV, respectively (DP1--4 in Fig.\ 1(d)). These fourfold bulk DPs are protected by non-symmorphic symmetry ($\widetilde{M}_{\hat{z}}:\{m_{001}|1/2\ 1/2\ 0\}$) combined by time-reversal and inversion ($P\Theta$) symmetry \cite{Young_Kane} (see also SM). Usually, TDS surface states are expected to manifest as higher dimensional objects, like e.g.\ Fermi arcs \cite{HfSiS}. In stark contrast, the surface states reported here are Dirac-like and emerge at the $\bar{X}$ point of the surface BZ at an energy position that is located between the bulk Dirac crossings. These characteristics are reminiscent to the surface states of topological insulators.
%  It is tempting to try to explain this salient finding by drawing on the non-symmorphic symmetries in conjunction to recent works on 2D semimetals \cite{Young_Kane} where the space group of ZrGeTe is the same as the one discussed there. If one regards the (001) surface as a purely 2D system, then, by directly applying the theory of Ref.\ \cite{Young_Kane}, one expects non-symmorphic symmetry protected Dirac crossings at the X$_1$, X$_2$, and M points of the 2D BZ (here the ZrGeTe surface) which would be pinned there by TRS. 
%%  Coming back to our surface state, this picture would imply that there should be dirac crossings at X' and M. 
%  Although we do calculate and also observe the crossings at the X points, our calculations do not yield any gapless state at M. Even so, one would still need to understand the origin (if not coincidental) of the surface states at the specific energies and momenta and why such gapless states are not present on the surface of other family members. For this purpose, here we follow the usual route of a topological analysis of the bulk states. 

In ZrGeTe there exists a momentum dependent gap between bands forming DP1, 2 and DP3, 4 [Fig.\ 1(d)] throughout the whole bulk BZ. Our calculations yield that the bands forming the DP3 and DP4 [see Fig.\ 1(d)] are inverted leading to a double band inversion at R (see SM). Each band inversion leads to the set of weak topological indices ($\nu_0$;\,$\nu_1 \, \nu_2 \, \nu_3$)=($0$;\,$1\, 1\, 0$) which would give rise to gapless Dirac surface states at the equivalent $\bar{X}$ points of the (001) surface. Since there are two band inversions in total, the overall parity does not change and this leads to vanish overall $Z_2$ indices, unless an additional symmetry can protect the surface states thus leading to a TCI phase \cite{TCI,TCI2}. 

Our analysis (see SM) gives that each band inversion is characterized by a different eigenvalue of the non-symmorphic symmetry operator that protects each bulk DP. Observing that $\widetilde{M}_{\hat{z}}=P C_{2\hat{z}}$, with $C_{2\hat{z}}:\{2_{001}|1/2\ 1/2\ 0\}$, we have that at the (001) surface where $P$ is broken, $\widetilde{M}_{\hat{z}}$ is broken as well, but $C_{2\hat{z}}$ remains intact. The latter is a crystalline symmetry that protects the resulting gapless surface states, leading to similar phenomenology as in TCIs. In fact, taking into account all remaining symmetries of the (001) surface, the resulting $k\cdot p$ model near $\bar{X}$ is similar to the archetypical one for TCIs \cite{TCI2}, leading to a pair of anisotropic Weyl-like surface states at $\bar{X}$. The surface states given by our \textit{ab initio} slab calculations disperse linearly (quadratically) along the $\bar{\Gamma}-\bar{X}$ ($\bar{M}-\bar{X}$) direction thus complying with this picture. Moreover, we find that the expected surface mass term \cite{TCI2} in the case of ZrGeTe is small so that the two surface states are near degenerate, thus giving rise to the bright spot seen in our ARPES data at the crossing point as shown in Fig.\ 3(f).

 Hence, our analysis proposes that ZrGeTe is the first topological material where TDS and TCI phases coexist under the protection of space-group symmetry. Notably, and in contrast to the usual TCIs, ZrGeTe is the first paradigm of a TCI that stems from multiple band inversions at the same time-reversal invariant momentum of the bulk BZ. A plausible reason why gapless surface states occur in ZrGeTe but not in other MSiX materials, are the different atomic size of Ge in contrast to Si and/or enhanced spin-orbit coupling that may allow for band inversions to occur.
%For example, our calculations indicate that some of the nodal lines along the X-R directions observed in ZrSiTe \cite{ZrSiTe} are split in ZrGeTe, which implies that SOC in ZrGeTe is stronger.
In this respect, ZrGeTe appears as a rare example of a naturally fine tuned system so that it can exhibit such rich phenomenology.

%now we discuss some of our observations.1. We experimentally discovered the presence of topological nodal-line phase in ZrGeTe.2. Importantly, we discovered the gapless topological surface state at around X point of the Brillouin zone where the Dirac point is located at about 400 meV below the chemical potential.3. Based on the theoretical calculations we find the possible origins of such gapless surface states in ZrGeTe which are its greater atomic size of Ge and/or enhanced spin-orbit coupling as compared to the other MSiX materials. These may allow band inversion to occur in ZrGeTe. 4. Furthermore, we explained this gapless surface states in terms of symmetry arguments. We find that the combined action of two screw axis operations, which together produce the non-symmorphic rotation and protects the gapless surface states. 5. More interestingly, we find that the non-symmorphic rotation further protects the weak topological insulator states Z2 invariants (0; 110). Therefore, our system potentially provides the so far unexplored correlation between topological nodal-line and topological insulator phases

 In conclusion, we performed systematic investigations of the ZrGeTe system. Our experimental ARPES data and \textit{ab initio} calculations reveal the existence of two-fold symmetry protected Dirac-like gapless surface states.
% multiple Fermi pockets and identifies their origin. Most importantly, we revealed the existence of non-symmorphic symmetry protected gapless surface states in the electronic structure of ZrGeTe. 
Moreover, our calculations reveal the presence of a nodal-line state along the M-$\Gamma$-M direction. Our study thus  highlights the existence of topological Dirac semimetal and topological insulator states in a single material, and opens up new prospects for studying the as yet unexplored interplay of these exotic quantum states.\\~\\ 
Acknowledgment\\
M.N.\ is supported by the start-up fund from University of Central Florida.
	 T.D.\ is supported by NSF IR/D program. 
	 A.A., P.M., and P.M.O.\ acknowledge support from the Swedish Research Council (VR), the R\"ontgen-{\AA}ngstr\"om Cluster, and the Swedish National Infrastructure for Computing (SNIC).
	 I.B.\ acknowledges the support of the NSF GRFP.
	 %D.K.\ was supported by the National Science Centre (Poland) under research grant 2015/18/A/ST3/00057.
	 Work at Princeton University is supported by the Emergent Phenomena in Quantum Systems Initiative of the Gordon and Betty Moore Foundation under Grant No.\ GBMF4547 (M.Z.H.) and by the National Science Foundation, Division of Materials Research, under Grants No.\ NSF-DMR-1507585 and No.\ NSF-DMR-1006492.
	 We thank Sung-Kwan Mo and Jonathan Denlinger for beamline assistance at the LBNL.

$*$ Correspondence and requests for materials should be addressed to M.N. (Email: Madhab.Neupane@ucf.edu).

\clearpage

\setcounter{figure}{0}
\renewcommand{\figurename}{\textbf{Supplementary figure}}
\begin{quote}
	\centering
\textbf{Supplementary Materials}
\end{quote}
\bigskip
\bigskip
%\*Correspondence and requests for materials should be addressed to M.N. (Email: Madhab.Neupane@ucf.edu).
\begin{figure*}[h!]
	\centering
	\includegraphics[width=16.5cm]{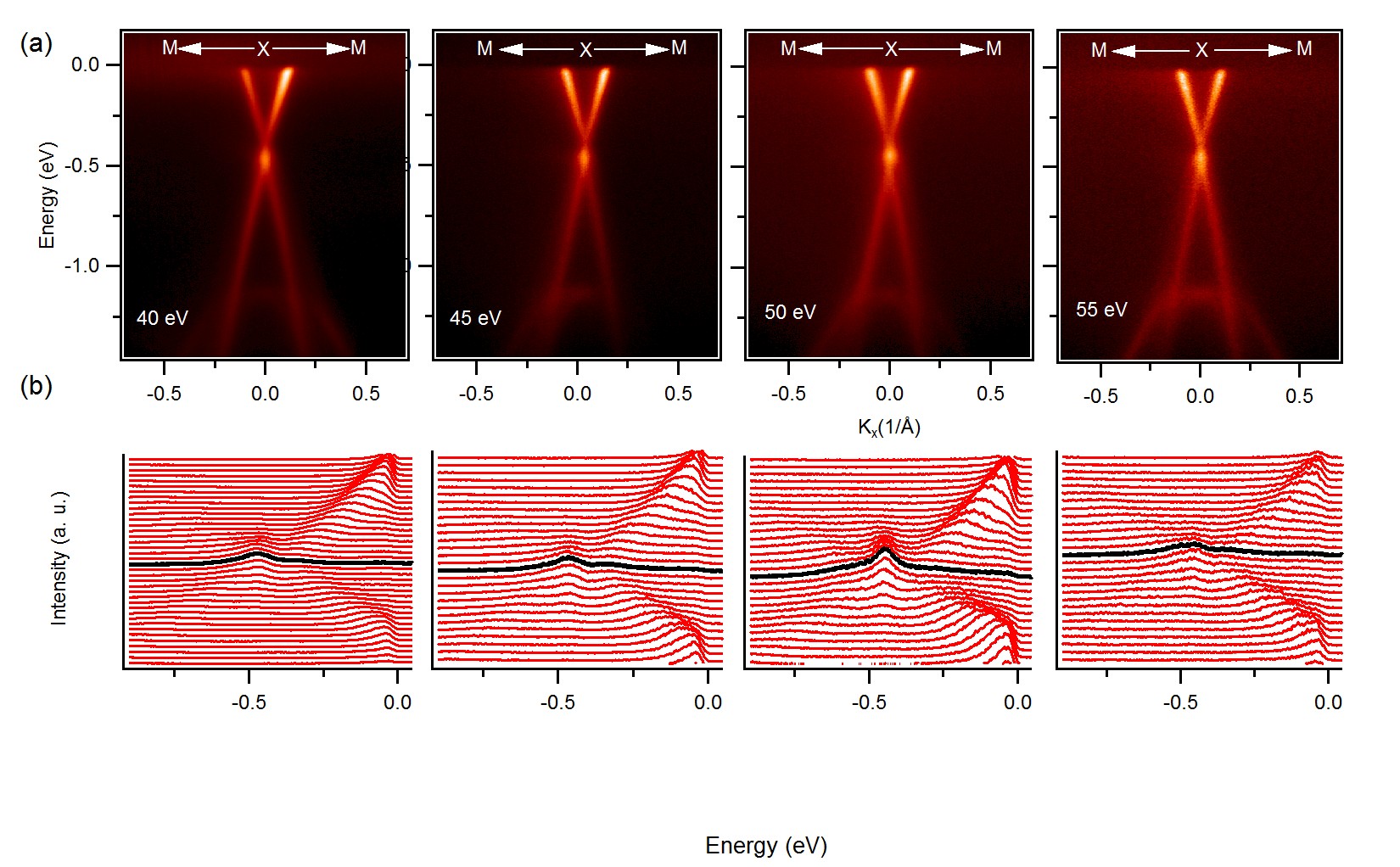}
	\caption{Observation of the gapless surface state. 
		{(a)} Dispersion maps along the M-X-M direction.
		(b) The energy distribution curves correspond to the panels shown under (a).}
	\label{figS1}
\end{figure*}

\begin{figure*}
	\centering
	\includegraphics[width=16.5cm]{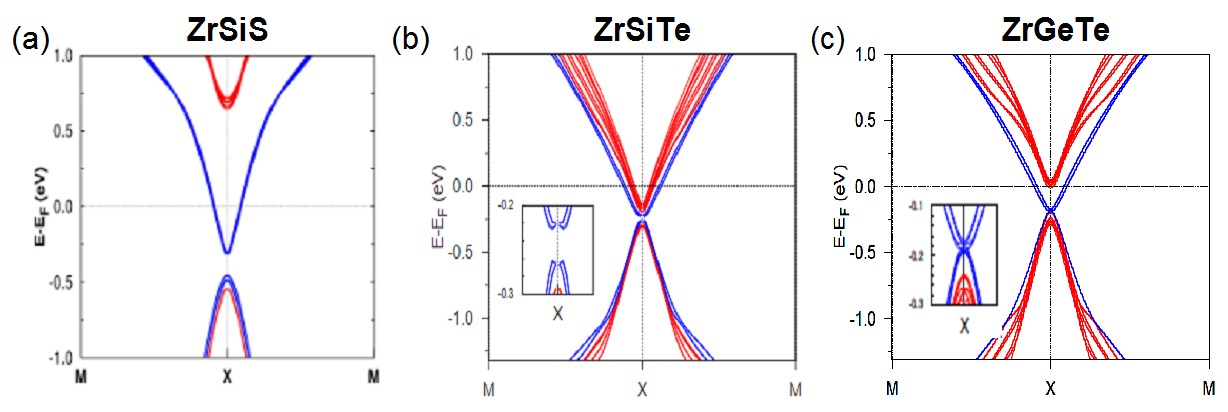}
	\caption{
		{Comparison of \textit{ab initio} calculated surface states between ZrSiS, ZrSiTe, and ZrGeTe}.  
		{(a)-(c)} Calculated dispersions along the M-X-M 
		direction of ZrSiS, ZrSiTe, and ZrGeTe, respectively.  Blue colored curves depict the surface states. Note that the surface states appear twice due to the presence of top and bottom surfaces in the calculations.}
	\label{figS2}
\end{figure*}
\bigskip

\begin{raggedleft}
	\textbf{Method.}
\end{raggedleft}

\begin{raggedleft}
	\textit{Crystal growth and transport characterization.}
\end{raggedleft}

Single crystals of ZrGeTe were grown by chemical vapor transport method using iodine as transporting agent. The crystals had a form of massive platelets with dimensions up to 10 $\times$ 5 $\times$ 3 mm\textsuperscript{3}. They were stable against air and moisture. The chemical composition was proven by energy-dispersive X-ray analysis using a FEI scanning electron microscope equipped with an EDAX Genesis XM4 spectrometer. The crystals were found to be single-phased and homogeneous with the desired equiatomic stoichiometry.

The crystal structure was examined at room temperature on a Kuma-Diffraction KM4 four-circle X-ray diffractometer equipped with a CCD camera using Mo K$\alpha$ radiation. The experiment confirmed the tetragonal ZrSiS-type structure (space group (SG) No.\ 129, 
$P4/nmm$) and yielded the lattice parameters: $a = 3.867(8)$ \AA \space and $c = 8.579(6)$ {\AA} (see an example in Fig.\ 1c). Electrical resistivity measurements were carried out in the temperature range $2-300$ K employing a conventional four-point ac technique implemented in a Quantum Design PPMS platform. The electrical contacts were made using silver epoxy paste. The electrical resistivity was measured with electrical current flowing within the tetragonal $a-b$ plane in zero magnetic field and in an external magnetic field of 9 T applied along the $c$-axis.
\\~\\
\begin{raggedleft}
	\textit{ARPES measurements.}
\end{raggedleft}

The synchrotron based experiments were performed at the ALS beamlines 10.0.1 and 4.0.3 equipped with R4000 and R8000 hemispherical electron analyzer, respectively. For the synchrotron measurements the energy resolution was set to better than 20 meV and the angular resolution was set to better than 0.2$^{\circ}$. Samples were cleaved \textit{in-situ} and measured at a temperature of $15-20$ K.
\\~\\
\begin{raggedleft}
	\textit{Theoretical calculations.}
\end{raggedleft}

\begin{figure*}
	\includegraphics[width=0.95\textwidth]{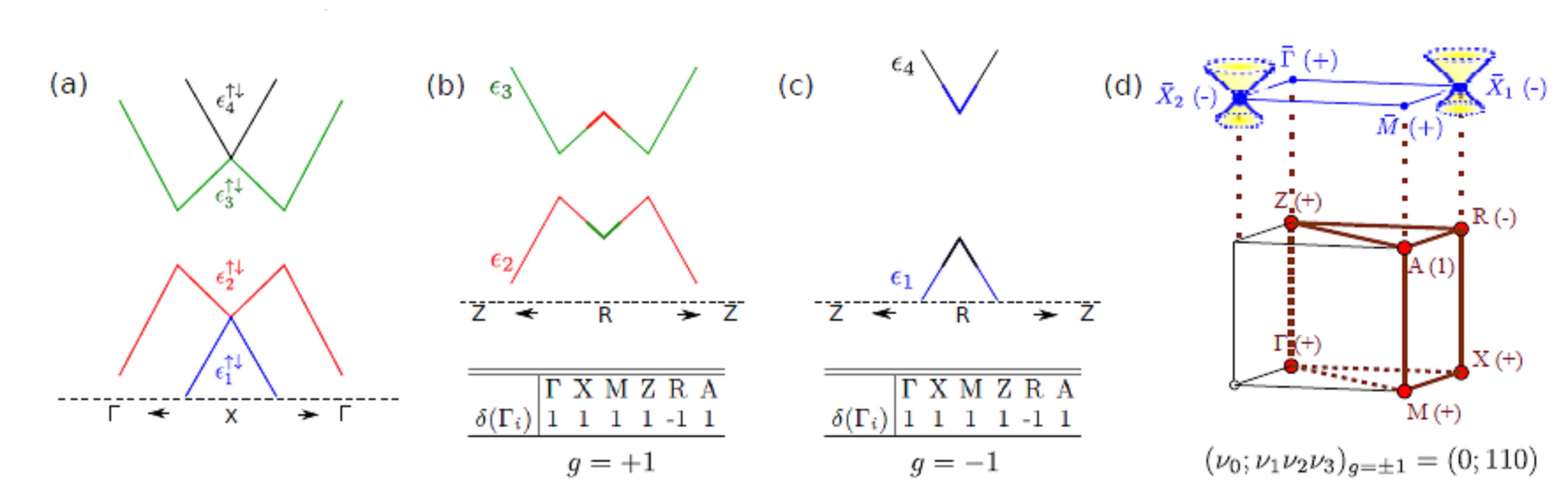}
	\caption{(a) Schematics of the eight bands forming the two Dirac points around the X (R) point. Here $\uparrow$($\downarrow$) denotes a pseudospin in the presence of SOC corresponding to the $P\Theta$ eigenvalue $+i$($-i$). (b),(c) Our \textit{ab initio} calculations yield two band inversions at $R$ (and the equivalent TRIM) in bulk ZrGeTe. The calculated products of parity eigenvalues at the non-equivalent TRIMs are shown in the tables. As we write below each table, each band inversion corresponds to a different eigenvalue of the $\widetilde{M}_{\hat{z}}$ nonsymmorphic symmetry operator, as discussed in the sections below. (d) Bottom: Schematics of the first octant of the BZ and non-equivalent high-symmetry points (TRIMs). In parenthesis the parity of $\delta(\Gamma_i)$ is shown. Top: Brillouin zone of the (110) surface that is probed by our ARPES measurements. The bulk inversions induce two Dirac surface states at the $\bar{X}_1$ and $\bar{X}_2$, respectively, which we observed by ARPES. For each surface state we have the same set of weak $Z_2$ indices but in different $\widetilde{M}_{\hat{z}}$ symmetry channels, a situation reminiscent to  crystalline topological insulators.}
	\label{supfig1}
\end{figure*}
The electronic structure calculations were carried out using the Vienna \textit{Ab-initio} Simulation Package (VASP) \cite{Kress_1S, VASPS}, and the generalized gradient approximation (GGA) used as the DFT exchange-correlation functional  \cite{Blochl_1S}. Projector augmented-wave pseudopotentials \cite{GGA_1S} were used with an energy cutoff of 500 eV for the plane-wave basis, which was  sufficient to converge the total energy for a given \textit{k}-point sampling. In order to simulate surface effects, we used 1 $\times$ 5 $\times$ 1 supercell for the (001) surface, with a vacuum thickness larger than 19  \AA. The Brillouin zone integrations were performed on a special \textit{k}-point mesh generated by 25 $\times$  25 $\times$  25 and 40 $\times$  40 $\times$  1 $\Gamma$-centered Monkhorst-Pack $k$-point grid for the bulk and surface calculations, respectively. The spin-orbit coupling (SOC) was included self-consistently in the electronic structure calculations. The electronic minimization algorithm used for static total-energy calculations was a blocked Davidson algorithm.

\bigskip

\begin{raggedleft}
	\textbf{Observation of gapless surface states.}
\end{raggedleft}
\\~\\
MSiX-type (where M = Zr, Hf and X = S, Se, and Te) materials are extensively studied both theoretically and experimentally by various groups and it has been unambiguously discovered that these materials possess a nodal-line phase and the surface states along the M-X-M direction. However, the observed surface states were found to be gapped. Our study solves this puzzle and reports the first direct observation of gapless surface states in ZrGeTe both experimentally and theoretically. For this purpose we present measured dispersion maps and their energy distribution curves along the M-X-M direction. In both figures we clearly observe the gapless surface states (see Fig.\ \ref{figS1}). Figure \ref{figS2} shows the comparative calculated dispersion maps of ZrSiS \cite{Neupane_5S}, ZrSiTe \cite{ZrSiXS}, and ZrGeTe, where the gapless surface state is found in ZrGeTe and the gapped surface states are observed in ZrSiS and ZrSiTe.
%This bulk DP might interfere with the observed surface state and thus lead to the increased intensity. In such a case, one would like to separate between the bulk and surface contributions. Since the surface state is 2D, this should remain dispersionless along kz, as is beautifully shown in Fig. 3f in main text, but the bulk contribution shouldn?t. However, since this bulk DP probably forms a nodal line along kz its dispersion along kz is subtle. Indeed, this is what we get from the calculation. Nevertheless, the calculation gives that the bulk DP (here DP4, for clarity) slightly moves towards more negative energies as we go from kz near R (surface) towards kz near X (bulk) which I believe (please correct me here) corresponds to going from left to right in Fig. 3f.From this trend, one would expect the bright spot to slightly move downwards in energy as we go further right in Fig.3(f) in main text which is fairly consistent with our experimental data (see Supplementary Figure 2). 
%Similar mass accumulation in ARPES intensity also observed in other studied and explained in terms of short life time of surface quasiparticles and bulk-surface resonance effect \cite{Trang, Donath}.
%\bigskip

\begin{raggedleft}
	\textbf{Topological analysis.}
\end{raggedleft}
%\\~\\

\begin{raggedleft}
	\textit{Band inversions and calculated $Z_2$ indices.}
\end{raggedleft}
%\\~\\

We begin by noticing that in ZrGeTe there is a momentum-dependent bulk gap that separates the bands forming the DPs at $X$ and $R$ throughout the whole Brillouin zone (BZ) and that the gapless surface states at $\bar{X}$ reside in this gap. Our calculated band structure near $X$ and $R$ has roughly similar shape as the one we show schematically in Fig.\ \ref{supfig1}a). The topological content of an insulator with time-reversal and inversion symmetry in 3D can be characterized by four Z$_2$ indices $(\nu_0;\nu_1\,\nu_2\,\nu_3)$, the first characterizing strong TIs ($\nu_0$) and the other three characterizing weak TIs when being nonzero. The inversion symmetry simplifies greatly their calculation and the indices only depend on the sign of the parity products of occupied bands at the time-reversal invariant momenta (TRIMs) of the Brillouin zone \cite{FuKanePRBS}:
\begin{eqnarray}\label{z2eq1}
\delta(\Gamma_i)=\prod_{m=1}^N\xi_{2m}(\Gamma_i)
\end{eqnarray}
where $\xi_{2m}(\Gamma_i)$ is the corresponding parity eigenvalue ($\pm 1$) at a TRIM point ($\Gamma_i$) of the $m$-th occupied band. The index $m$ runs through all occupied bands ($N$) and $2m$ means that we only account for one partner of each Kramers pair. Following the standard procedure \cite{FuKanePRBS}, for SG No.\ 129 the four Z$_2$ indices are given by the products,
\begin{eqnarray}
(-1)^{\nu_0}&=&\delta(\Gamma)\cdot\delta(M)\cdot\delta(Z)\cdot\delta(A) , \\
(-1)^{\nu_1}&=&\delta(X)\cdot\delta(M)\cdot\delta(R)\cdot\delta(A)=(-1)^{\nu_2} , \\
(-1)^{\nu_3}&=&\delta(Z)\cdot\delta(A) ,
\end{eqnarray}
where, by symmetry, $X_1$ is equivalent with $X_2$ (we denote them both as X) and the same holds for $R_1$ and $R_2$ (we denote them both as R).

One can find the eigenvalue parities $\xi(\Gamma_i)$ by directly calculating the hybridized wavefunctions at the TRIMs and then feed them back into Eq.\ (\ref{z2eq1}). However, this procedure is often too cumbersome and usually low energy $k\cdot p$ effective models are employed to provide the information on the parities, instead. Here we follow an alternative procedure which permits us to directly use our \textit{ab initio} calculated band structure, overcoming the need for explicit wavefunction calculations.
This we achieve by looking at the sign of the mass gap between every two consecutive bands at the TRIM points which in particle-hole symmetric systems is unambiguously related to the parity eigenvalues \cite{FuKanePRBS}. In particular, in particle-hole symmetric systems, a negative sign of the gap between two bands at a TRIM point means that a band inversion has occurred between these two bands at that point and concomitantly gives $\delta(\Gamma_i)=-1$ for the valence band. In order to apply these ideas to a realistic bandstructure, we first establish particle-hole symmetry between each pair of bands taking advantage of the adiabatic continuity property of the topological state, i.e., without ever closing the gap. 
%A detailed account of the procedure will be given elsewhere.

For ZrGeTe, we find that band inversions between bands 2, 3 and 1, 4 take place at the R point of the BZ (see Fig.\ \ref{supfig1}). Each band inversion is characterized by the calculated parity products and Z$_2$ indices that are shown in the Tables of Fig. \ref{supfig1}b) and Fig.\ \ref{supfig1}c). 
%As discussed in the previous section, each band inversion is characterized by an eigenvalue, $g=\pm 1$, of $\widetilde{M}_{\hat{z}}$ as shown in Figures \ref{supfig1}a) and \ref{supfig1}b). 
As a result each band inversion at R yields the set of weak topological indices ($\nu_0$;$\nu_1 \,\nu_2 \,\nu_3$)=($0$;$1 \,1 \,0$) which would give rise to gapless Dirac surface states at the equivalent $\bar{X}$ points of the (001) surface as shown in Fig.\ \ref{supfig1}d). However, since there are two band inversions in total, the overall parity does not change and this leads to vanish overall $Z_2$ indices. Concomitantly, topological (gapless) surface states cannot arise unless there is an extra symmetry protection. In the following sections we argue that in fact we are dealing with a new type of  crystalline topological insulating phase that supports this kind of surface states.
\\~\\
\begin{raggedleft}
	\textit{Analysis of inverted Dirac points in the bulk.}
\end{raggedleft}

The crystal structure of ZrGeTe, belonging to SG No.\ 129, possesses three nonsymmorphic symmetries which, as pointed out recently \cite{Young_KaneS}, protect the bulk Dirac points thus giving rise to a nonsymmorphic symmetry-protected Dirac semimetal. Here we focus on the glide mirror symmetry $\{g|{\bf t}\}=\{m_{001}|\frac{1}{2} \frac{1}{2} 0\}=\widetilde{M}_{\hat{z}}$ which maps the $X_1$ point onto the $X_2$ point of the BZ. As Young and Kane \cite{Young_KaneS} showed, under time-reversal symmetry (TRS) the bulk Dirac points at X$_{1(2)}$,R$_{1(2)}$ (DP$i$, $i=1,3$ and $i=2,4$ respectively) are protected by two nonsymmorphic symmetries, $\{g|{\bf t}\}$. These consist of a glide mirror $\widetilde{M}_{\hat{z}}$ and a screw rotation $\widetilde{C}_{2\hat{x}(2\hat{y})}$. Here we focus on $X_1$ but similar conclusions can be drawn for $X_2$ by taking $\widetilde{C}_{2\hat{x}}\leftrightarrow\widetilde{C}_{2\hat{y}}$. Also, we will refer to $X_1$, but the same analysis applies to $R_1$ as well since the lines $\Gamma-X_1$ and $\Gamma-R_1$ have the same symmetries and satisfy $g{\bf k}={\bf k}$, $e^{i{\bf G t}}=-1$, where ${\bf t}$ is a half-translation and ${\bf G}$ is a reciprocal lattice wave vector. We will differentiate between X and R further below.

\begin{figure*}
	\includegraphics[width=0.5\textwidth]{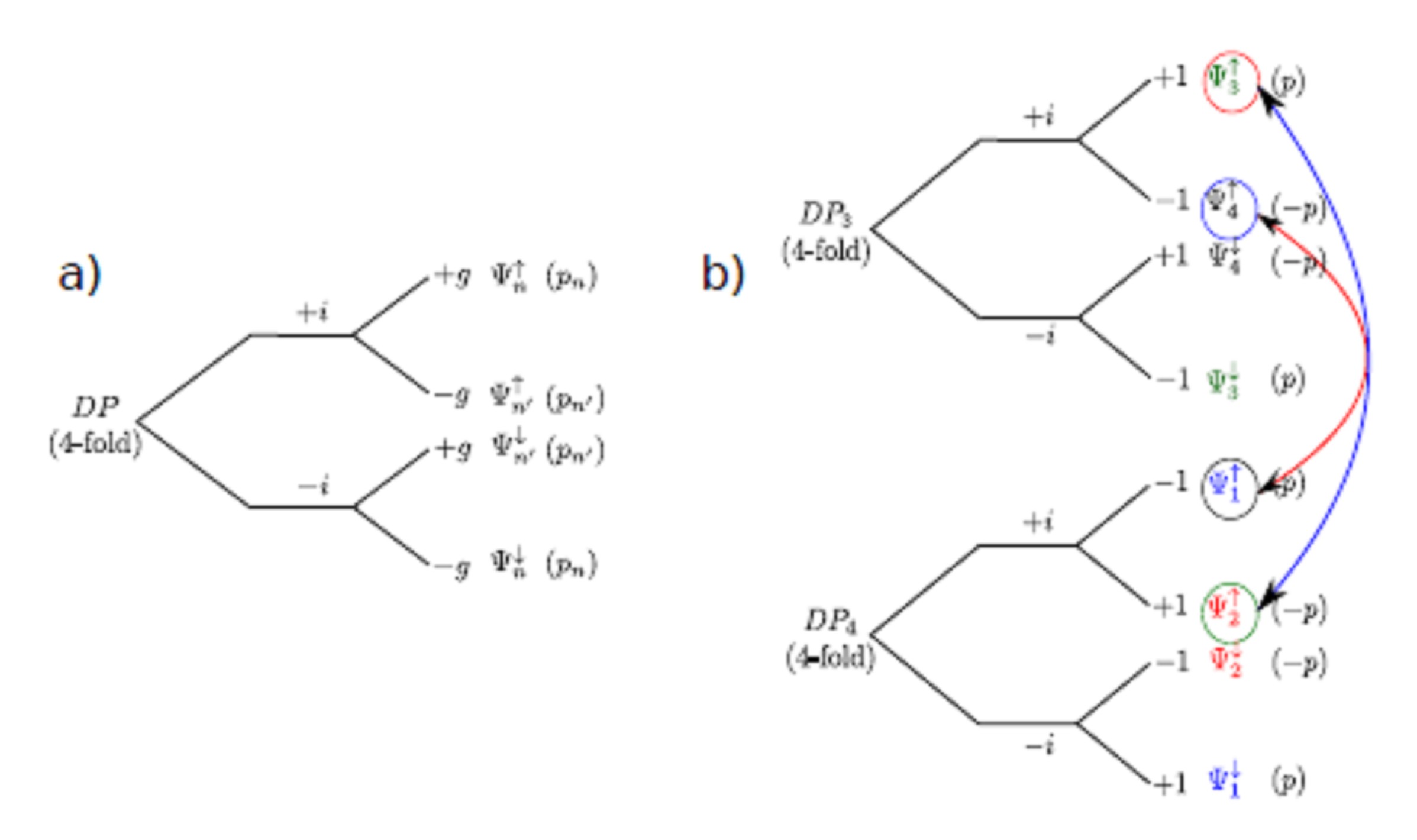}
	\caption{(a) Schematic of the generic DP structure discussed in the text. The eigenvalues of $\widetilde{M}_{\hat{z}}$ are denoted with $g$, those of P with $p$, and $n,n'$ are band indices. (b) Schematic of the two DPs at and of a double band inversion process at the R point(s).}
	\label{fig2sup}
\end{figure*}

In the presence of SOC, under TR ($\Theta$) and parity ($P$) symmetry we have $(P\Theta)^2=-1$. At the X point, $\widetilde{M}_{\hat{z}}^2=1$, $\Theta^2=-1$, $P^2=1$ and we have that $[\Theta,P]=[\Theta,\widetilde{M}_{\hat{z}}]=0$, $\{P,\widetilde{M}_{\hat{z}}\}=0$ thus guaranteeing a fourfold degeneracy which constitutes a Dirac point since at $\Gamma$, $\widetilde{M}_{\hat{z}}^2=-1$, i.e. changes sign \cite{Young_Kane}. The DP is pinned at $X_1$ due to TRS. The above commutation and anti-commutation relations imply that Kramers pairs have opposite $\widetilde{M}_{\hat{z}}$ eigenvalues ($g$), i.e., $\widetilde{M}_{\hat{z}}(P\Theta\Psi_g)=-g(P\Theta\Psi_g)$, where $\widetilde{M}_{\hat{z}}\Psi_g=g\Psi_g$ and of course the same parity eigenvalues ($p$), i.e., $P(P\Theta\Psi_p)=p(P\Theta\Psi_p)$, where $P\Psi_p=p\Psi_p$. 

This analysis  leads to the generic form of a DP as shown in Fig.\ \ref{fig2sup}a), where $\uparrow,\downarrow$ are pseudospins that correspond to $\pm i$ eigenvalues of $P\Theta$, $n,n'$ are band indices and each parenthesis indicates the parity eigenvalue, $p_{n(n')}$. At the X (R) points, each of the DP's obeys this structure while the band structure consists of a V ($\mathcal{M}$) shaped band touching with a W ($\Lambda$) shaped band near the DP that is above (below) the Fermi level, as seen for example in the schematic of Fig.\ \ref{fig2sup}a). 

We argue that for a band inversion (and a concomitant change in parity) to occur, the exchanged Kramers partners must share the same eigenvalue under $\widetilde{M}_{\hat{z}}$ but opposite parities. Since after the band inversion(s) take place the resulting bulk Dirac points need still be  protected by the same symmetries, their structure in terms of Fig.\ \ref{fig2sup}a) should not change. This constrain requires that a double band inversion must simultaneously occur, or in other words, at the TRIM 
%(time reversal invariant momentum) 
the two DPs must be inverted. As an example, Fig.\ \ref{fig2sup}b) shows how a band inversion between bands 2 ($\mathcal{M}$-shaped) and 3 (W-shaped) can take place denoted by the blue line. In the same figure, denoted by red line, we draw an inversion between bands 1 ($\Lambda$-shaped) and 4 (V-shaped). This picture is verified by our DFT calculations and the calculations presented in the previous section.
\\~\\
\begin{raggedleft}
	\textit{Discussion of surface states.}
\end{raggedleft}

Our calculations for the $Z_2$ invariants indicates that such double band inversions take place in ZrGeTe at the R points (e.g.\ between DP$3$ and DP$4$). Due to the constraint that the inverted DPs need still to form bulk Dirac points (the bulk remains a Dirac semimetal) the parity inversions happen within the same $\widetilde{M}_{\hat{z}}$ symmetric channel, i.e., each exchanged Kramers pair carries the same $\widetilde{M}_{\hat{z}}$ eigenvalue, therefore for $g=\pm 1$ we have the same set of weak $Z_2$ indices, $(0;1\,1\,0)$ characterizing each inversion (as indicated below each table in Fig.\ \ref{supfig1}a) and Fig.\ref{supfig1}b)). This situation is reminiscent of what happens in TCIs. There, multiple band inversions at bulk TRIM points can project onto the same surface TRIM point, therefore annihilate each other unless they are protected by an additional crystalline symmetry that survives at the surface \cite{TCIS}.

%\begin{figure}[h!]
%\includegraphics[width=0.65\textwidth]{fig3sup}
%\caption{(a) Schematic of the eight bands forming the two Dirac points around the X (R) point. Here $\uparrow$($\downarrow$) denotes a pseudospin in the presence of SOC corresponding to $P\Theta$ eigenvalue $+i$($-i$). b) Schematic of the generic DP structure discussed in the text. The eigenvalues of $\widetilde{M}_{\hat{z}}$ are denoted with $g$, those of P with $p$ and $n,n'$ are band indices. c) Schematic of the two DPs at and of a double band inversion process d) inverted DP structure at the R point(s).}
%\label{fig3sup}
%\end{figure}

As already discussed, the fourfold bulk DPs are protected by nonsymmorphic symmetries ($\widetilde{M}_{\hat{z}}$) combined by $P\Theta$ symmetry. At the surface, $P$ and $\widetilde{M}_{\hat{z}}$ are broken leaving only $\Theta$ intact. In fact, the nonsymmorphic symmetries $\{2_{100}|\frac{1}{2} 0 0\}=\widetilde{C}_{2\hat{x}}$ and $\{2_{010}|0 \frac{1}{2} 0\}=\widetilde{C}_{2\hat{y}}$ are broken at the surface, as well. However, $M_x=\widetilde{M}_{\hat{z}}\widetilde{C}_{2\hat{y}}$, $M_y=\widetilde{M}_{\hat{z}}\widetilde{C}_{2\hat{x}}$ and $C_{2\hat{z}}=M_x M_y$ (=$-\widetilde{C}_{2\hat{x}} \widetilde{C}_{2\hat{y}}$) survive at the surface. It is possible to decompose $\widetilde{M}_{\hat{z}}$ in terms of parity and rotation symmetries: $\widetilde{M}_{\hat{z}}=P C_{2\hat{z}}$. Seen from this viewpoint, the surface breaks parity (and therefore $\widetilde{M}_{\hat{z}}$) but $\Theta$ \textit{and} $C_{2\hat{z}}$ remain intact. The latter is a crystalline symmetry that protects the resulting gapless surface states, leading to similar phenomenology as in TCIs. From the previous relations and $\{C_{2\hat{z}},P\}=0$, $\{\widetilde{M}_{\hat{z}},P\}=0$, $\{C_{2\hat{z}},\widetilde{M}_{\hat{z}}\}=0$, we have that states with the same $\widetilde{M}_{\hat{z}}$ eigenvalue ($g_z$) but opposite parities have the same $C_{2\hat{z}}$ eigenvalue. This means that in the band inverted DPs, the parity inversion is accompanied by an inversion in the $C_{2\hat{z}}$ eigenvalues. 
%This can be seen schematically in Fig. \ref{fig3sup}a) where we explicitly follow the internal degree of freedom due to $g_{2z}$. 

In total, the remaining symmetries at the (001) surface are $\Theta$, $C_{2\hat{z}}$ and $M_x$, $M_y$. These lead to a $k\cdot p$ theory near the $\bar{X}$ point of the surface BZ that is similar to the one first pointed out for TCIs in Ref.\ \cite{TCI2S} and thus similar results apply here, as well. As discussed in Ref.\ \cite{TCI2S}, there is a single surface mass term allowed which splits the 4-fold degeneracy leading to two anisotropic Weyl-like gapless surface states at the $\bar{X}$ point. In the case of ZrGeTe, our slab calculations yield that this mass term turns out to be very small so that the surface states appear almost 4-fold at $\bar{X}$, although their true nature is Weyl-like. In addition, the same $k\cdot p$ analysis predicts that in principle these Weyl-like dispersions can be linear along one high-symmetry direction and quadratic along another one. Our \textit{ab initio} calculations are in total agreement with the predicted anisotropy and we can explain the linear (quadratic) dispersions along the $\bar{\Gamma}-\bar{X}-\bar{\Gamma}$ ($\bar{M}-\bar{X}-\bar{M}$), shown in Fig.\ 3(e) (Fig.\ 3(g)) of the main text, as signatures of the topological crystalline origin of the observed surface states.

%
%
%\begin{figure*}
%\centering
%\includegraphics[width=16.5cm]{Fig5}
%\caption{\textbf{Absence of bulk $\Gamma$-pocket and coupled temperature dependence of the in-gap states.}
%(\textbf{a}) Hybridized bands in the vicinity of the Fermi level along the $\Gamma-{\textrm{X}}$ momentum space cut. A bandgap of about 15 meV is obtained in our calculation (see supplementary information for details)
%(\textbf{b}) Orbital decomposed band structure near Fermi level along the $\Gamma-{\textrm{X}}$ line. The size of blue and yellow spheres are proportional to the weight of Sm 5$d$ and 4$f$ orbital, respectively.
%(\textbf{c}) A comparison of integrated EDCs featuring the $\Gamma$ pocket and the ${\textrm{X}}$ point bands and the low-lying in-gap states. }
%A gap value of about 14 meV is observed in both cases\end{figure*}

\end{document}